\documentclass[aps,prl,twocolumn,groupedaddress,showpacs]{revtex4}

\usepackage{epsfig}
\begin{document}

\title{Magnonic Analogue Black/White Hole Horizon in Superfluid $^3$He-B: experiment}

\author{M. \v{C}love\v{c}ko}
\author{E. Ga\v{z}o}
\author{P. Skyba}
\email[Electronic Address: ]{skyba@saske.sk}

\affiliation{{\mbox Institute  of Experimental Physics, SAS} and P. J. \v{S}af\'arik
University Ko\v{s}ice, Watsonova 47, 04001 Ko\v{s}ice, Slovakia.}

\date{\today}

\begin{abstract}
We provide experimental details of the first experiment made in zero temperature limit
($\sim$ 600\,$\mu$K) studying  the magnonic black/white hole horizon analogue using
absolutely pure physical system based on the spin superfluidity in superfluid $^3$He-B.
We show that spin precession waves propagating on the background of the spin
super-currents in a channel between two Bose-Einstein condensates of magnons in form of
homogeneously precessing domains mimic the properties of the black/white horizon. Once
the white hole horizon is formed and blocks the propagation of the spin-precession waves
between two domains, we observed an amplification effect, i.e. when the energy of the
spin precession waves reflected from the horizon is higher than the energy of the excited
spin precession waves before horizon was formed. Moreover, the estimated temperature of
the spontaneous Hawking radiation in this model system is about four orders of magnitude
lower than the system's background temperature what makes it a promising tool to study
the effect of spontaneous Hawking radiation.

\end{abstract}


\maketitle

\section{INTRODUCTION}

One of the most exciting outcomes of the quantum field theory in curved space-time is the
Hawking`s famous prediction that the gravitational black holes are not stable objects,
they are spontaneously evaporating by radiating particles due to quantum fluctuations on
the event horizon \cite{hawking}. As this radiation has a thermal spectrum, the black
holes have  non-zero temperature (so-called the Hawking temperature), the magnitude of
which is inversely proportional to the black hole mass. However, a smallness of the
Hawking temperature (for solar mass black hole $\sim$ 60\,nK) in comparison with the
temperature of the cosmic microwave background radiation ($\sim$ 3\,K) makes the direct
experimental measurement of the Hawking temperature using current technologies
impossible. Solving conceptual issues  with theoretical treatment of the Hawking
radiation - so-called trans-Planckian problem, W. Unruh showed that the behaviors of
classical and quantum fields in curved space-times are more-less analogical, and
suggested to test the validity of Hawking's prediction using experimentally created
black-hole-horizon. In particular, by measuring a thermal spectrum of sound waves emitted
from the sonic horizon in transonic fluid flow \cite{unruh1,visser}. This suggestion has
triggered and motivated as theoretical so experimental search for physical systems which
could serve as a model tool mimicking the properties of the black-hole horizon.

The physical principle of an experimental black/white hole (or event horizon) formation
is based on the controlled transition between subcritical and supercritical (and
vice-versa) regimes of the propagation of a particular perturbation on the background of
the flow allowing to distribute the perturbation over physical system. A black-hole
horizon for perturbation in a flowing medium is formed by a transition from subluminal to
superluminal flow, such that perturbation spreading along the flow and passing from the
subluminal to superluminal region cannot return back to the subluminal region. On the
other hand, a white-hole horizon is a region where the flow changes from superluminal to
subluminal. In this case, a perturbation traveling against the flow from the subluminal
region cannot penetrate the superluminal part.

Except above mentioned sound waves in trans-sonic fluid flow, the portfolio of such
systems were extended by: atomic Bose-Einstein condensate \cite{garay}, light in
dispersive media \cite{light}, the surface waves on flowing fluid \cite{ralf,rousseaux1,
rousseaux2,silke}, hydraulic jumps in flowing liquids \cite{volovik, jannes}, light in
optical fibre \cite{philbin} and ultra-cold fermions \cite{BEC2}. Therefore, in light of
this, the Hawking radiation can be viewed as universal phenomenon of a dynamic nature in
the various physical systems having capability, at certain conditions, to form a boundary
- an event horizon, and the fundamental dynamical property of any event horizon analogue
is a spontaneous emission of thermal Hawking radiation, the temperature of which depends
on the magnitude of the velocity gradient on the horizon \cite{unruh1,ralf1}
\begin{equation}
T=\frac{\hbar}{2 \pi k_B}\frac{\partial v^{r}}{\partial r} \sim 10^{-12} \frac{\partial v^{r}}{\partial r}.
\end{equation}\label{equ1}
It turns out however, that this temperature is typically a several orders of magnitude
lower than the background temperature of the physical systems used as experimental tool
modeling properties of the black/white hole horizon. Solution to this problem seems to be
to find a physical system, the background temperature of which would approaching the
absolute zero temperature, and simultaneously would have the properties allowing to model
the black/white horizon \cite{garay,BEC2,BEC3,magnonic}.

In this article we provide all experimental details and discuss the  results  of the
first experiment made in zero temperature limit ($\sim$ 600\,$\mu$K) studying the
magnonic black/white hole horizon analogue using absolutely pure physical system based on
the spin superfluidity in superfluid $^3$He-B \cite{spinhole}. Main experimental results
were published in \cite{prl2019}.

\section{IDEA OF EXPERIMENT}

\begin{figure}[htb]
\begin{center}
\epsfig{figure=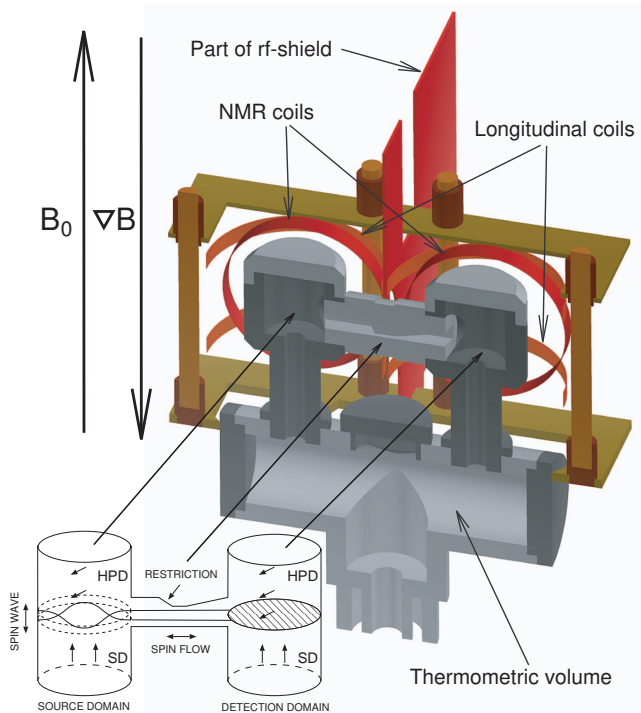,width=85mm} \caption{(Color on line) Schematic 3-D cross section
of the experimental cell and concept of the experiment. The NMR coils and the longitudinal coils
are Helmholtz type coils and served for the HPD and the spin-precession waves excitation, respectively. }
\label{fig1}
\end{center}
\end{figure}

First let us explain the concept of the experiment. The schematic 3-D cross section of
the experimental chamber and concept of the experiment is shown in Fig. \ref{fig1}. The
experimental chamber with superfluid $^3$He-B consists of two cylindrical cells mutually
connected by a channel. The chamber is placed in steady magnetic field $B_0$ and magnetic
field gradient $\nabla$B, both oriented along $z$-axis.

Using cw-NMR technique, we created a Bose-Einstein condensate of magnons in a form of the
homogeneously precessing domain (HPD) in both cylindrical cells. The HPD is a dynamic
spins structure formed in a part of the cell placed in lower magnetic field which, with
an aid of the dipole-dipole interaction, coherently precess around a steady magnetic
field at the angular frequency $\omega_{rf}$ \cite{hpd1,fomin}. Within the rest of cell,
the spins are co-directional with the steady magnetic field and do not precess. These two
regions (domains) are separated by a planar domain wall, the position of which is
determined by the Larmor resonance condition $\omega_{rf}= \gamma (B_0 + \nabla B.z)$,
where $\gamma$ is the gyromagnetic ratio of the $^3$He nuclei.

To model black/white hole horizon in superfluid $^3$He-B, we used two fundamental
physical properties of the HPD: the spin (magnonic) superfluidity and the presence of the
HPD's collective oscillation modes in a form of the spin precession waves (see below).
The spin superfluidity allows to create and manipulate the spin flow, i.e. the spin
super-currents flowing between precessing domains as a consequence of the phase
difference $\Delta \alpha_{rf}$ between the phases of spin precession in individual
domains \cite{moscow}. The spin precession waves serve as a probe testing formation and
presence of a black/white hole horizon inside the channel: channel has restriction
allowing to reach the different regimes of the velocity of the spin super-current flow
with respect to the group velocity of the travelling spin-precession waves. The spin
super-currents affect the transfer of the spin precession waves i.e. axial oscillation
modes of the domain wall excited by longitudinal coil in one of the HPD (source domain)
and detected in the second, receiving HPD (detection domain) as a function of the phase
difference in the spin precession of HPDs, i.e. as a function of the spin flow velocity.
This is similar to the experiment suggested by R. Sch\"{u}tzhold and W. Unruh \cite{ralf}
and later performed by G. Rousseaux et al. \cite{rousseaux1,rousseaux2} and by S.
Weinfurtner et al. \cite{silke}. In the presented experiment, the magnetic field gradient
$\nabla$B plays the role of gravitational acceleration, the spin super-currents represent
the water flow, and the spin-precession waves correspond to the gravity waves on the
water's surface.

\section{EXPERIMENT}

As mentioned above, the experiment was performed in the experimental cell, the 3-D
cross-section of which is shown in Fig. \ref{fig1}. The inner diameters of both cylinders
were 6\,mm and heights were 7\,mm. The channel connecting two cylinders was 13\,mm long
and in order to create a spatial gradient in spin flow velocity $u$, the channel had an
asymmetric reduction in its center. The length of sharpest restriction was 0.5\,mm and
after restriction the channel dimensions were reduced. The reduced channel was 2\,mm long
and its cross-section had dimension of 0.4 $\times$ 3\,mm$^2$ (see Fig. \ref{fig1}). The
experimental cell was mounted on a diffusion-welded copper nuclear stage and cooled down
\cite{stage}. The cell was filled with $^3$He at temperature of $\sim$ 1\,K, and then
pressurized to 3\,bars. Subsequent demagnetizations of the copper nuclear stage allowed
to cool $^3$He into superfluid state, down to temperature $\sim$ 0.5 T$_c$, at which the
measurements were performed. Temperature of $^3$He was measured by means of a powdered
Pt-NMR thermometer placed in bottom part of thermometric volume (Pt-NMR thermometer is
not shown in Fig. \ref{fig1}). Temperature was calibrated against T$_c$ - the transition
temperature of $^3$He into superfluid state.

Two HPDs were simultaneously and independently generated in both towers using two
programmable function rf-generators working in the phase-locked mode at angular frequency
$\omega_{rf}=2\pi.462\times10^3$\,rad/sec and with zero phase difference
$\Delta\alpha_{rf}$ adjusted between excitation voltage signals. The induced NMR signals
were amplified by pre-amplifiers and measured by two rf-lock-in amplifiers, each
controlled by its own generator. As the position of the domain walls in both towers
follows the plane, where the Larmor resonance condition is satisfied, i.e. $\omega_{rf} =
\gamma (B_0 + \nabla B.z)$, the position of both domain walls (i.e. also the size of the
HPDs)  can easily be controlled and adjusted by steady magnetic field $B_0$ (a current
source supplying the NMR magnet).

\begin{figure}[t!]
\begin{center}
\epsfig{figure=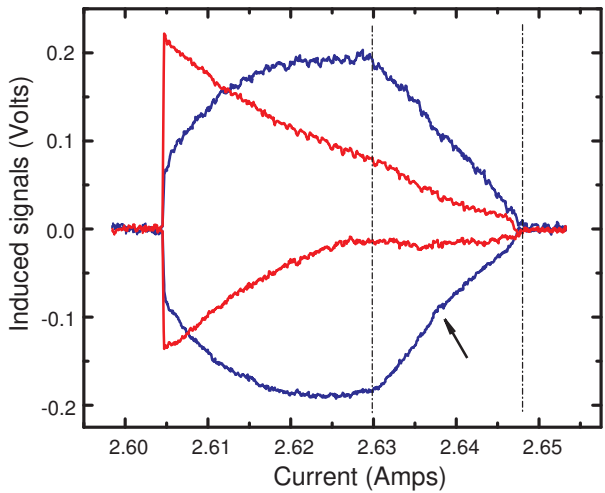,width=80mm} \caption{(Color online) Example of the voltage signals corresponding to
absorption (red) and dispersion (blue) of two HPDs generated at frequency 462\,kHz in both cylinders as
function of the magnetic field (current). The dot-dashed lines indicate the beginning and the end of the
cylindrical cell. Arrow indicates the change in the slope. Signals from one HPD were
artificially inverted in order to clarify measured dependencies.} \label{fig2}
\end{center}
\end{figure}

Figure \ref{fig2} shows the absorption (in red) and dispersion (in blue) NMR signals from
two simultaneously generated HPDs as a function of the current supplying the NMR magnet.
While absorption signals are proportional to the energy dissipation, the amplitude of the
dispersion signals correspond to the HPD volume. There are  three relaxation processes
leading to the  power dissipation  in   the  HPD.  A total  HPD power dissipation  in a
cylindrical experimental  cell of  radius $R$  with $z$-axis  of the cell orientated
along the magnetic field and its gradient can be expressed as \cite{jltp1997}
\begin{equation}
\dot{Q} = \sigma \frac{\chi }{\gamma^2} \frac{D_{\perp}\omega_{rf}^2}
{\lambda_F}\cdot S + 2\pi w_sR\cdot \mathcal{L} + \frac{5\chi}{6}\tilde \tau_{LT}
S \frac{\omega_{rf}^2 (\nabla \omega)^2}{\gamma^2} \,\mathcal{L}^3.
\end{equation}\label{equ2}

The  first  term  describes the relaxation  of the two-domain structure  by  the  spin
diffusion  across the domain  wall  of  cross-section   $S$   with $D_{\perp}$ being the
transverse component of the spin-diffusion tensor for gradients along  the $z$-direction.
Here $\lambda_F$ is the domain wall thickness,  $\sigma$  is  the domain wall
shape-dependent  constant of the order of unity and $\chi$ is the temperature dependent
susceptibility of the BW state with Fermi-liquid corrections. The second term represents
a relaxation process caused by the distortion of the bulk order parameter near the walls
of the experimental cell, where $w_{s}$  is the constant characterizing  the surface
relaxation rate \cite{surf}. The  third  term  refers   to  the  relaxation  through  the
intrinsic  Leggett-Takagi  mechanism  with  $\tilde \tau_{LT}$ being the effective
Leggett-Takagi  relaxation  time, where $\nabla \omega = \gamma \nabla B$ and
$\mathcal{L}$ is the HPD length in cylindrical cell \cite{lt}.

In Fig. \ref{fig2} the dot-dashed lines mark the beginning and the end of the
experimental cell. While sweeping magnetic field down, both HPDs begin to be formed on
the top of the cell, when the Larmor resonance condition is satisfied there. On the
further reduction of the magnetic field, the Larmor resonance condition moves along $z$ -
axis of the cell, and so do the domain walls of both HPDs. The dispersion signals
increase linearly as the volume of the each HPD grows in cylindrical cell. Penetration of
the domain walls into the channel is associated with the change in slopes of measured
signals, mainly dispersion ones. Asymmetry in absorption signals from HPDs is caused by a
spin current flowing between the HPD due to a non-zero phase shift between rf-excitation
fields used for HPDs generation. In fact, what determines the magnitude of the spin flow
between domains is the phase difference between rf-fields (i.e. the electric currents)
generated by the excitation coils and not the phase difference in excitation voltage
signals provided by rf-generators. As the resonance circuits are not perfectly matched,
there is a few degrees phase shift between rf-fields, no matter that a zero phase
difference between voltage excitation signals from locked generators was adjusted.
Resulting spin flow transfers the energy between domains that leads to asymmetry in
measured signals. When both HPDs fulfil entire cell, the dispersion signals saturate.
Further field reduction causes the increase of the energy dissipation due to the
Leggett-Takagi relaxation mechanism, until the rf-fields (with aid of the spin
super-currents) can not cover the energy losses and both HPDs ``decease". It is worth to
note that when the domain wall is placed inside the cell (the case of this experiment),
the dominant process of the energy dissipation is the spin diffusion.

Once two HPDs were generated, the position of the domain wall has to be adjusted into
channel. Although, the penetration of the domain walls into channel manifests itself as a
change in the slope of the dispersion signals (see Fig. \ref{fig2}), we used another
method based on generation of the collective oscillations modes of the HPD-SD structure
(see below), and application of this method allowed us to adjust the domain wall position
in the channel with resolution of a few tens of $\mu$m (depending on the value of the
magnetic field gradient).

\begin{figure}[h!]
\begin{center}
\epsfig{figure=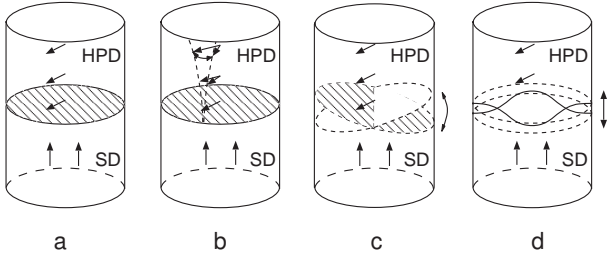,width=80mm} \caption{(Color online) A schematic visualization of various
kinds of the collective oscillation modes - spin precession waves of the HPD-SD structure: (a) the HPD-SD stable
state, (b) the torsion oscillation mode, (c) the planar mode, and (d) the first axial surface mode.
The last mode - the first axial mode was used as a probe to adjust the domain wall position into channel
and to test the formation of the event horizon in superfluid $^3$He-B. } \label{fig3}
\end{center}
\end{figure}

Figure \ref{fig3} shows a schematic visualization of the various kinds of collective
oscillations mode - spin precession waves of the HPD-SD spin structure, i.e. the various
forms of the small oscillations of the phase of the spin precession in the HPD. The
physics and the measurements methods of these modes are very well understood and
developed, respectively
\cite{Fomin87,torsional,moscow1,vovka,euro,waves,prl2008,matoprb2012}. The dispersion
relation for the spin-precession waves travelling along the HPD-SD interface placed far
from horizontal wall of the cell was derived in form \cite{matoprb2012}
\begin{equation}
\Omega^2 = \frac{3}{4}\frac{c_1 \gamma \nabla B}{\omega_{rf}}\sqrt{
\frac{4}{\sqrt{15}}\gamma B_{rf} \omega_{rf} + \frac{(5c_L^2+3c_T^2)\,k^2_T}{3}},
\end{equation}\label{equ3}
where $k_T = \xi_{m,i}/R$  is the transverse wave vector, $R$ is the cell radius,
$\xi_{m,\,i}$, are zeroes of the derivative of the Bessel function, $c_{L}$  and  $c_{T}$
denote the spin wave velocities with respect to the longitudinal or
transversal direction to the magnetic field direction, respectively, and $c_1 =
\sqrt{2(5c_T^2-c_L^2)/3}$.

In presented experiment we used the first axial surface mode as a probe testing the
formation of the event horizon in channel. The procedure is as follows: as the first
step, for given value of the magnetic field gradient, the domain wall was set at the
position above the channel using steady magnetic field $B_0$, that is at the position
before the change in the slope of the dispersion signals occur. Then we added small
harmonically oscillating longitudinal magnetic field generated by one coil, and using the
frequency sweep of this longitudinal field we found the first axial oscillation mode of
the domain wall.

\begin{figure}[htb]
\begin{center}
\epsfig{figure=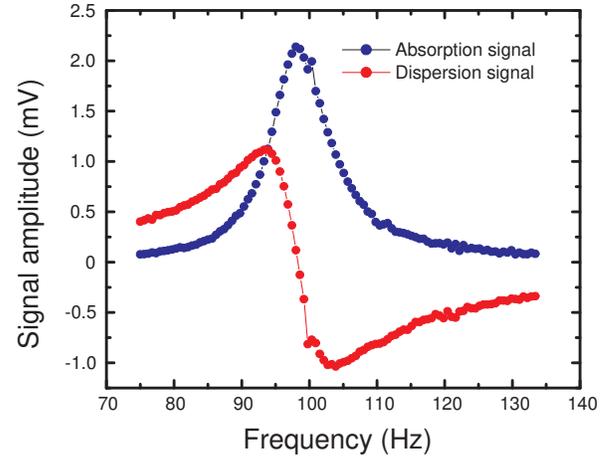,width=80mm} \caption{(Color online) Absorption and dispersion resonance
characteristics of the first axial oscillation mode of the HPD-SD spin structure.} \label{fig4}
\end{center}
\end{figure}

Figure \ref{fig4} shows the resonance characteristics of a spin precession wave (the
first axial mode) excited by the longitudinal coil in one HPD, when the domain walls are
placed above the channel at the constant field gradient $\nabla B$. Then, by sweeping the
position of domain wall (i.e. sweeping the magnetic field $B_0$) along the experimental
cell and simultaneously continuously exciting the longitudinal coil on the resonant
frequency of the spin wave, the position of the domain wall in channel can be determined
from the amplitude of measured resonance signal.

\begin{figure}[h!]
\begin{center}
\epsfig{figure=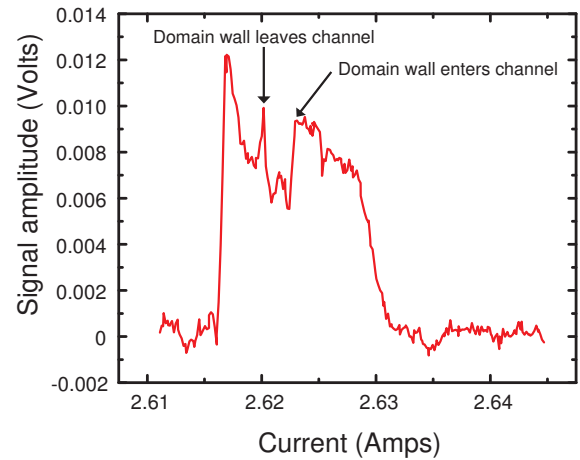,width=80mm} \caption{(Color online) Amplitude of the spin precession wave signal as a
function of the domain wall position (i.e. magnetic field B$_0$). } \label{fig5}
\end{center}
\end{figure}

Figure \ref{fig5} presents such dependence of the amplitude of the spin wave resonance
signal as a function of the domain wall position, i.e. the value of the current supplying
NMR magnet. There is obvious abrupt decrease of the signal, when the domain wall
penetrates into channel being followed by sudden signal increase, when the domain wall
leaves the channel. Latter change serves as a reference level to adjust the position of
the domain wall in the channel, as the channel base is flat. In fact, reaching the value
of the current (field $B_0$), at which the domain wall left the channel, we reversed the
current and returned the domain wall back to the channel by adding 4-5 steps of the
current source. The precision of the domain wall adjustment in channel  is given by
$\Delta B_0/ \nabla B $, where $\Delta B_0 = 0.76\,\mu$T is the field step controlled by
the current source \cite{source}, and for $\nabla B$=15 mT/m giving a spatial precision
of $\sim$ 50 $\mu$m. The estimated length of the precessing layer  in the channel $L$ is
$\sim$ 150\,$\mu$m $\pm$ 50 $\mu$m. For comparison, the domain wall thickness $(\lambda_F
= c_L^{2/3}/(\gamma \nabla B \omega_{rf})^{1/3})$ for the above mentioned experimental
conditions is $\lambda_F \sim$ 0.34\,mm.

The flow of the spin super-currents $J_\perp$ between precessing domains can be created
by the phase gradient of the spin precession $\nabla\alpha_{rf}$ \cite{Fomin87}
\begin{equation}
 J_\perp=-\frac{\chi}{\gamma}\left[(1-\cos\beta)^2c^2_L+(1-\cos^2\beta)c^2_T\right]\nabla\alpha_{rf} \; ,
\end{equation}
where $\beta$ is the spin deflection angle. The spin flow $ J_\perp$ between precessing
domains can be established in both directions depending on the sign of the phase
difference $\Delta\alpha_{rf}$, while the spin flow velocity $u$ depends on the magnitude
of $\nabla\alpha_{rf}$ as
\begin{equation}\label{equ5}
u=\frac{(5c_L^2-c_T^2)\,\nabla \alpha_{rf}}{2 \omega_{rf}}.
\end{equation}
The group velocity of the propagating spin precession waves $c$ is expressed as
\begin{equation}\label{equ6}
c^2=\frac{(5c_L^2+3c_T^2)\,\gamma \nabla B L}{4\omega_{rf}}.
\end{equation}
Here $c_L$ and $c_T$ denote already mentioned the longitudinal and transverse spin wave
velocities with respect to the magnetic field orientation, respectively, and $L$ is the
HPD length in the channel. We shall assume that $\nabla \alpha_{rf} $ is localized on the
length of the sharpest restriction in the channel of the order $dl$ = 0.5\,mm, therefore
$\nabla \alpha_{rf} \sim \Delta \alpha_{rf} /dl$.

\begin{figure}[h!]
\begin{center}
\epsfig{figure=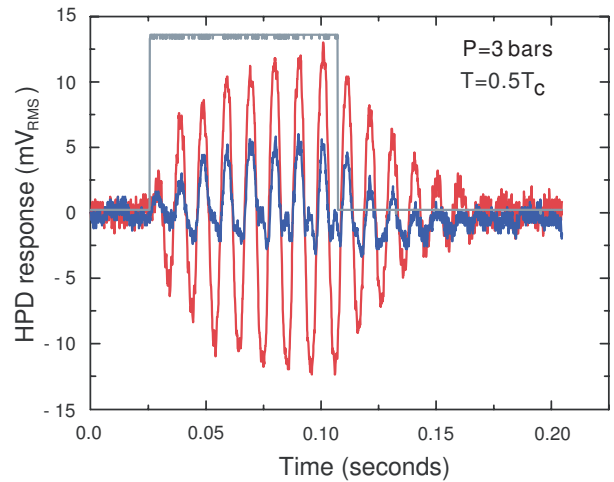,width=80mm} \caption{(Color on line) Voltage signals
corresponding to spin-precession waves: the source domain (red), the detection domain (blue).
The rectangular signal shows the time window when 8 sinusoidal excitation pulses were applied in
order to excite the spin-precession waves.} \label{fig6}
\end{center}
\end{figure}

After setting the position of the domain wall in the channel and adjusting the phase
difference $\Delta\alpha_{rf}$ i.e. adjusting the amplitude and direction of the spin
flow velocity $u$, we excited the spin-precession waves. The spin-precession waves in the
source domain were generated by applying 8 sinusoidal pulses at their resonance frequency
using separate generator. The low frequency signals representing spin precession waves
from the both, the source and detection HPDs for a particular value of $\Delta
\alpha_{rf}$ were extracted using demodulation technique i.e. by means of rf-detector and
a low-frequency filter. These low frequency signals were subsequently captured and stored
by a digital oscilloscope for the data analysis. Figure \ref{fig6} shows examples of the
time evolution of the corresponding signals of the excited spin-precession waves in the
source domain and incoming waves in the second, detection domain. The excitation pulse
builds-up the spin precession waves.  When the pulse is over, the spin precession waves
decay due to dissipation mechanisms. We analysed just these free decay signals by the
methods of the spectral analysis as a function of the phase difference $\Delta
\alpha_{rf}$.

\begin{figure}[htb]
\begin{center}
\epsfig{figure=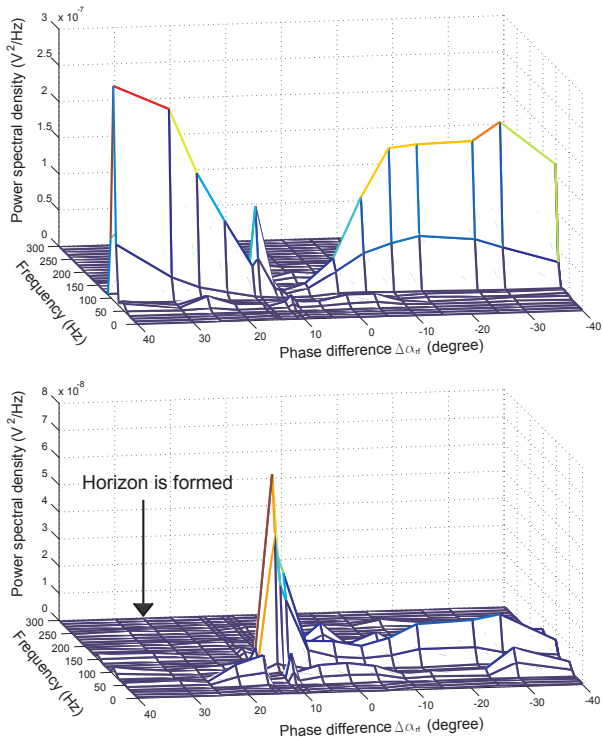,width=90mm} \caption{(Color on line) The power spectral density of
the free decay signals as function of the phase difference measured from the source domain
(upper) and the detection domain (bottom). For details see text.} \label{fig7}
\end{center}
\end{figure}

\section{Results and Discussion}

\begin{figure}[htb]
\begin{center}
\epsfig{figure=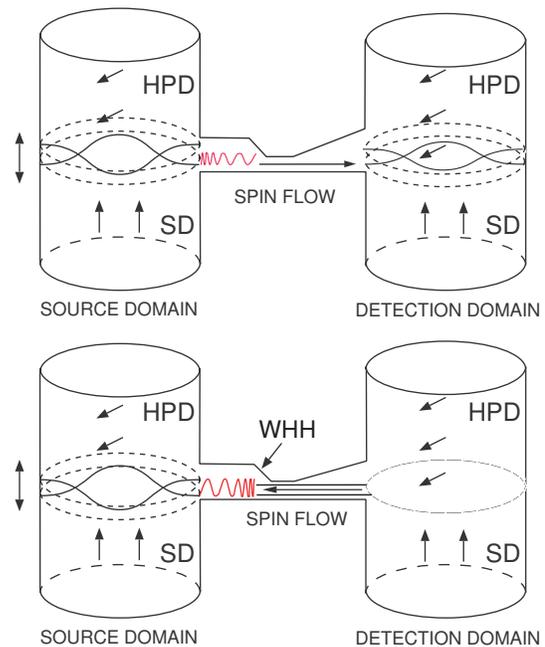,width=70mm}
\caption{(Color on line) A schematic illustration of the spin precession waves dynamics in channel.
Upper figure shows propagation of spin precession waves in direction of the spin super-currents flow.
Bottom figure illustrates a formation of the white hole horizon i.e. the case when spin precession waves
propagate against the spin super-currents flow. For details, see text.}
\label{fig8}
\end{center}
\end{figure}

Figure \ref{fig7}  shows the power spectral density (PSD) of the free decay signals
measured in the source (upper) and in the detection (lower) domains as a function of the
phase difference $\Delta \alpha_{rf}$ i.e. as a function of the direction and the
amplitude of the spin flow velocity. As one can see, there are a few remarkable
characteristics to be seen in these dependencies. Firstly, the upper figure shows the
strong PSD signals being detected in the source domain with an exception of a deep
minimum in region of $\Delta \alpha_{rf}$ corresponding to $\sim$ 10$^\circ$. Secondly,
the lower figure shows the weaker PSD signals of the incoming spin precession waves
detected in the detection domain in the range of negative values of $\Delta \alpha_{rf}$
up to 10$^\circ$, above which a peak signal was detected. Thirdly, no PSD signals (within
experimental resolution) were measured in the detection domain for values of $\Delta
\alpha_{rf} \gtrsim$ 20$^\circ$. We interpret observed dependencies as follows. A
schematic illustration of the spin precession waves dynamics in the channel are presented
in Fig. \ref{fig8}.

The negative values of the phase difference $\Delta \alpha_{rf}$ cause the spin
super-current to flow from the source domain towards the detection one (see Fig.
\ref{fig8}). Thus, the spin-precession waves excited in the source domain are dragged by
these spin-currents and travel downstream to the detection domain, where they are
detected (see Fig.\ref{fig7} bottom). The frequency of excited spin-precession waves corresponds to that of the
fundamental axial mode of an isolated domain (see Fig. \ref{fig4}). However, the
out-flowing or in-flowing spin super-current varies the conditions on the boundary of a
domain wall and thus modifies the frequency of its fundamental mode. Therefore, the
frequency of incoming spin-precession wave slightly differs from the resonance frequency
of the standing wave, and consequently, the amplitude of detected oscillations is lower
(reduction in the wave amplitude is also caused by the dissipation).

Presence of the deep signal minimum in the source domain and corresponding signal maximum in
the detection domain for $\Delta\alpha_{rf} \sim$ 10$^\circ$ - 15$^\circ$ we interpret
as a consequence of zero spin super-current between domains which leads to a resonance match between cavities.
We  remind that what determines the phase of the spin precession in HPDs is the phase of
the rf-field i.e. the phase of the electrical current flowing the resonance circuit and not the
phase of the excitation voltage provided by the generator (which we used as adjustable parameter).
A small mismatch between the real resonance frequency and the frequency of the voltage excitation
leads to a small phase shift between the excitation voltage and electrical current. This
explains why zero spin super-current flows between the domains at the phase difference
$\Delta\alpha_{rf} \sim$ 10$^\circ$ - 15$^\circ$. Now, from point of view of the axial oscillations,
two HPDs represent two resonance cavities connected by means of the channel. As the spin super-current
absences between HPD's, and the spin super-currents modify the boundary conditions for the axial mode,
this sets the condition for the resonance match between these cavities. Due to this, the whole
energy of the spin-precession waves excited in the source domain is almost immediately transferred
to and absorbed by the detection domain at once. This is detected as a missing signal
in the source domain, where the waves were excited (!), while the detection domain shows
a maximal PSD signal (see Fig. \ref{fig7}).

\begin{figure}[h!]
\begin{center}
\epsfig{figure=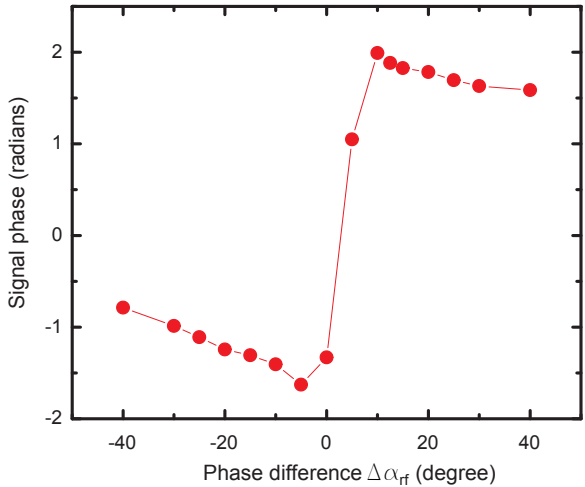,width=80mm}
\caption{(Color on line) Dependence of the phase of the free decay signals measured from
the source domain as a function  the phase difference $\Delta \alpha_{rf}$ showing
a gradual phase shift by 180$^{\circ}$.}
\label{fig9}
\end{center}
\end{figure}

For phase difference $\Delta \alpha_{rf}\,>$ 10$^\circ$ the flow direction of the spin
super-currents is reversed, i.e. the spin super-currents flow from the detection domain
towards the source domain and emitted the spin-precession waves propagate against this
flow. The change in the flow direction of the spin super-currents is also seen on the
phase of the decay signals from the source domain as a gradual phase shift by
~180$^\circ$ (see Fig. \ref{fig9} ). This is because the frequency of the spin precession
waves crosses from the values slightly lower to the values slightly higher (or vice
versa) than the resonance frequency of the spin-precession waves of the isolated HPD.

As one can see from Fig. \ref{fig7}, there are no signals detected in the detection
domain for $\Delta \alpha_{rf} \gtrsim$ 20$^\circ$. We interpret this as a formation of
the magnonic white hole horizon in the channel: the spin-precession waves sent from the
source domain towards to the detection domain are blocked in the channel by the flow of
spin super-currents, and never reach the detection domain (see Fig. \ref{fig8}). This
interpretation is supported by the calculation using above presented theoretical model:
the white hole horizon is formed in a place, where and when the  condition $c^2=u^2$ is
satisfied (see Eqs.~(\ref{equ5}) and (\ref{equ6})).

\begin{figure}[htb]
\begin{center}
\epsfig{figure=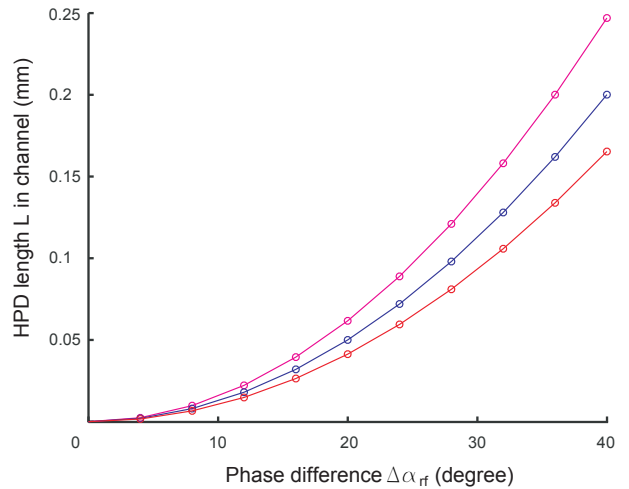,width=80mm}
\caption{(Color on line) Dependence of the HPD length $L$ in channel on the phase difference
$\Delta \alpha_{rf}$ that satisfy the equation  $c^2 = u^2$ (see Eqs. (\ref{equ5}) and (\ref{equ6})), that is
the condition for the white horizon formation in the channel. Curves are calculated for
different lengths of the sharpest restriction $dl$: 0.45\,mm (magenta), 0.5\,mm (blue) and 0.55\,mm (red).
} \label{fig10}
\end{center}
\end{figure}

Figure~\ref{fig10} shows a dependencies of the HPD length $L$ in the channel on the phase
difference $\Delta \alpha_{rf}$ that satisfy the ``event horizon" condition $c^2=u^2$
assuming that $\nabla \alpha_{rf}$  is localized and concentrated on the sharpest
restriction in the channel i.e. on the length of $dl$. Dependencies  were calculated for
three different values of $dl$ as stated in the caption of Fig. \ref{fig10}. Taking the
value of the phase difference $\Delta \alpha_{rf}$, at which a white horizon in the
channel is formed, i.e.  $\Delta \alpha_{rf} \geq$\,25$^{\circ}$ (see Fig. \ref{fig7}),
one can find that to satisfy theoretically predicted condition $c^2=u^2$ for horizon
formation, the corresponding length of the HPD $L$ in the channel should be of the order
of $L \sim$ 100\,$\mu$m, depending on the length of restriction. This value is in
reasonable agreement with that determined from experiment $L$ = 150 $\mu$m $\pm$ 50
$\mu$m.

\begin{figure}[htb]
\begin{center}
\epsfig{figure=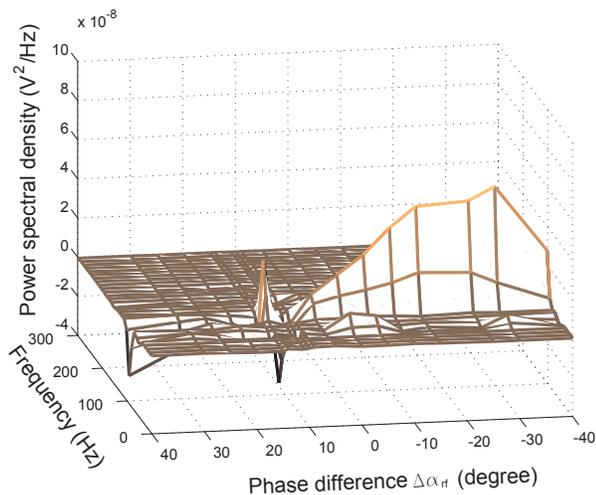,width=85mm} \caption{(Colour on line) The cross-correlation power spectral
density of the source and detector free decay signals as a function of
the phase difference $\Delta \alpha_{rf}$.} \label{fig11}
\end{center}
\end{figure}

Simultaneous measurement of both free decay signals of the spin precession waves from the
source and the detection domains allows a cross-correlation function between these two
signals to be determined. Figure \ref{fig11} shows the cross power spectral density
between these two signals as a function of the phase difference $\Delta \alpha_{rf}$ i.e.
as a function of the spin flow velocity $u$. As mentioned above, for values of $\Delta
\alpha_{rf} <$ 0$^\circ$, i.e. when the spin super-currents drag the excited spin
precession waves from the source domain towards to the detection domain, the decay
signals from both domains are correlated. For values of $\Delta \alpha_{rf}\,>$
0$^\circ$, reduction and subsequent reversion of the spin super-currents affects the
dynamics of the propagation of the spin precession waves that leads to the change in the
correlation between the decay signals detected in both domains. When the spin
super-currents approaches zero value, due to resonance match between domains, the energy
is transferred in both directions that is manifested as correlation/anti-correlation
peaks.  However, when the white horizon is formed ($\Delta \alpha_{rf} >$25$^\circ$), the
decay signals are anti-correlated. We may interpret this in a way that the rise of the
decay signal in source domain is paid by the spin flow flowing from detection domain
towards to the source domain - in agreement with theoretically predicted amplification of
the wave on the horizon paid by the energy of the flow \cite{unruh2}. This interpretation
is supported by the dependence presented in Fig. \ref{fig7} (upper dependence), where a
notable feature regarding to the absolute values is showed: when the spin super-current
flows from the detection domain to the source domain, the spin precession waves in the
source domain have a tendency to have a higher power spectral density amplitude than
those when the spin super-current flowing in opposite direction. However, to elucidate
the physical origin of the observed phenomena additional experiments have to be done.

\begin{figure}[th]
\begin{center}
\epsfig{figure=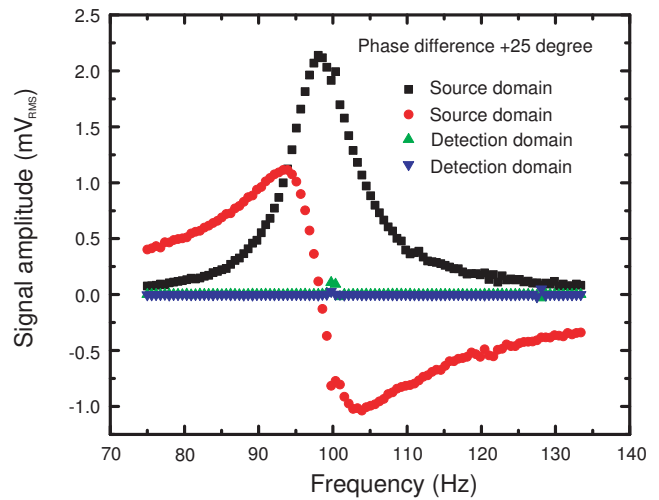,width=85mm} \caption{(Colour on line) Absorption and
dispersion signals of the spin-precession waves excited using continuous excitations
and measured in the source and the detection domains. The phase difference
$\Delta \alpha_{rf}$ between HPD rf-excitation voltages is equal to +25$^{\circ}$.} \label{fig12}
\end{center}
\end{figure}

Finally, to check the formation of the magnonic white hole horizon in channel we
performed measurements adjusting the phase difference between HPDs rf-excitation voltages
$\Delta \alpha_{rf}$ to be  +\,25$^{\circ}$, but instead using the pulse technique to
excite the spin precession waves we used the continuous excitations.  That is, we used a
full frequency sweep and simultaneously measured signals from both domains. Figure
\ref{fig12} shows absorption and dispersion signals from the source and detection
domains. The presence of the spin precession waves excited in the source domain is
manifested by strong signals. However, there are no signals detected in the detection
domain: as the spin precession waves excited in the source domain are propagating against
the spin super-current, they reach a horizon and they are reflected back never
penetrating into the detection domain.

\section{CONCLUSION}

In conclusion, we provided experimental details of an experiment in a limit of absolute zero temperature
probing the magnonic black/white hole horizon analogue in superfluid $^3$He-B. As an
experimental tool simulating the properties of the black/white horizon we used the spin
precession waves propagating on the background of the spin super-currents between two
Bose-Einstein condensates of magnons in form of homogeneously precessing domains.
We showed that by adjusting the external parameters of experiment, we can form a white hole
horizon inside the channel, and this horizon  blocks the propagation of the spin-precession waves between two
domains. Once the white hole horizon was formed, we observed an amplification effect,
when the energy of the reflected spin precession waves from the horizon is higher than
the energy of the spin precession waves excited before the horizon was formed. Even more,
the presented results manifest that the spin-precession waves propagating on the
background of the spin super-currents between two HPDs possess all physical features
needed to elucidate physics associated with the presence of the event horizons, e.g. to
test the spontaneous Hawking process. In fact, assuming that the spin super-currents
velocity of the order of $u \sim $ 1\,m/s varies on the length of $dl\sim 10^{-4}$\,m one
can estimate the temperature of the Hawking radiation in this system to be of the order
of 10 nK, what is a temperature only four orders of magnitude lower that the background
temperature, and this makes presented system a promising tool to investigate this
radiation.

\section{ACKNOWLEDGEMENT}

We acknowledge support from the European Microkelvin Platform (H2020 project 824109),
APVV-14-0605, VEGA-0128, CEX-Extrem SF of EU (ITMS 26220120047). We wish to acknowledge
the technical support provided by \v{S}. Bic\'{a}k and G. Prist\'{a}\v{s}. Financial
support provided by the US Steel Ko\v{s}ice s.r.o. is also very appreciated

\end{document}